\documentclass[preprint,showpacs,preprintnumbers,amsmath,amssymb]{revtex4}

\usepackage{graphicx}
\usepackage{dcolumn}
\usepackage{bm}
\usepackage{graphicx}
\usepackage{rotating}
\usepackage{dcolumn}
\usepackage{bm}

\begin{document}


\title{Vortices, circumfluence, symmetry groups and Darboux transformations of the (2+1)-dimensional Euler equation}
\author{S. Y. Lou$^{1,2}$, M. Jia$^{1}$, X. Y. Tang$^{1, 2}$ and F. Huang$^{1,2,3}$}
\affiliation{\small $^{1}$Department of Physics, Ningbo
University, Ningbo, 315211, China\\ \small $^{2}$Department of
Physics,
Shanghai Jiao Tong University, Shanghai, 200030, China\\
\small $^{3}$Department of Marine Meteorology, Ocean University of
China, Qingdao 266003, China}

\begin{abstract}
The Euler equation (EE) is one of the basic equations in many
physical fields such as fluids, plasmas, condensed matter,
astrophysics, oceanic and atmospheric dynamics. A symmetry group
theorem of the (2+1)-dimensional EE is obtained via a simple direct
method which is thus utilized to find \em exact analytical \rm
vortex and circumfluence solutions. A weak Darboux transformation
theorem of the (2+1)-dimensional EE can be obtained for \em
arbitrary spectral parameter \rm from the general symmetry group
theorem. \rm Possible applications of the vortex and circumfluence
solutions to tropical cyclones, especially Hurricane Katrina 2005,
are demonstrated.
\end{abstract}


\maketitle

\section{Introduction.}

There are various important open problems in fluid physics. One of
the most important problems is the existence and smoothness problem
of the Navier-Stokes (NS) equation. The NS equation has been
recognized as the basic equation and the very starting point of all
problems in fluid physics \cite{NS}. Due to its importance and
difficulty, it is listed as one of the millennium problems of the
21st century \cite{clay}.

One of the most significant recent developments related to the above
problem may be the discovery of Lax pairs of two- and three-
dimensional Euler equations (EEs) which are the limit cases of the
NS equation for a large Reynolds number \cite{Li,Li1}. Actually, for
the two-dimensional EE, the Lax pair given in \cite{Li,Li1} is weak
(see Remark 2 of the next section) while the Lax pairs of the
three-dimensional EE are strong (see Theorems 3 and 4 of \cite{LTJH}
which can be proved in a similar way as Theorem 1 of this paper).
Hence, the EEs are (weak) Lax integrable under the meaning that they
possess (weak) Lax pairs, and subsequently the NS equations with
large Reynolds number are singular perturbations of (weak)
Lax-integrable models.

The (3+1)-dimensional EE
\begin{eqnarray}
E_1&\equiv& \vec{\omega}_t+(\vec{u}\cdot\nabla)
\vec{\omega}-(\vec{\omega}\cdot\nabla) \vec{u}
=0,\label{E}\\
\vec{\omega}&=&\nabla \times \vec{u},\label{E1}\
\end{eqnarray}
with  $\vec{\nabla} \cdot \vec{u}=0$ is the original springboard for
investigating incompressible inviscid fluid. In Eqs. \eqref{E} and
\eqref{E1}, $\vec{\omega}\equiv\{\omega_1,\ \omega_2,\ \omega_3\}$
is the vorticity and $\vec{u}\equiv\{u_1,\ u_2,\ u_3\}$ is the
velocity of the fluid.

In (2+1)-dimensional case, the EE has the form of
\begin{eqnarray}
E\equiv\omega_t+[\psi,\omega]=0,\qquad  \omega&=&\psi_{xx}+\psi_{yy},\label{e}
\end{eqnarray}
where the velocity $\vec{u}=\{u_1,\ u_2\}$ is determined by the stream function $\psi$ through
\begin{eqnarray}
u_1=-\psi_{y},\ u_2=\psi_{x}\label{u12}
\end{eqnarray}
and the Jacobian operator (or, namely, commutator) $[A,\ B]$ is
defined as
\begin{eqnarray}
[A,\ B]\equiv A_xB_y-B_xA_y. \label{AB}
\end{eqnarray}

It is known that the EEs are important not only in fluid physics \cite{fluid} but also in many other physical
fields such as plasma physics \cite{plasma}, oceanography \cite{ocean}, atmospheric dynamics \cite{gas},
superfluid and superconductivity \cite{super}, cosmography and astrophysics \cite{astro}, statistical physics
\cite{sta}, field and particle physics\cite{particle} and condensed matter including Bose-Einstein condensation
\cite{bose}, crystal liquid \cite{crystal} and liquid metallic hydrogen \cite{H}, etc..

As a beginning point of various physical problems, the EEs have been
studied extensively and intensively, which is manifested by a large
number of related papers on EEs in the literature. For instance, a
lot of exact analytical solutions of the EEs have been presented,
some of which can be found in the classical book of H. Lamb
\cite{Lamb}. In \cite{Sov}, the authors studied the planar
rotational flows of an ideal fluid and the addressing method was
developed to obtain exact solutions of the EEs in \cite{Yu}. In
addition, some types of exact solutions were obtained via a
B\"acklund transformation in \cite{Jia}. However, rather few
 \em exact analytic \rm solutions of the EEs have been obtained from the (weak) Lax pair since it was revealed by
Charles Li \cite{Li} around five years ago. A special type of
Darboux transformation (DT) with zero spectral parameter for the
(2+1)-dimensional EE was shown in \cite{Li}, and some types of DTs
(or weak DTs) with \em nonzero \rm spectral parameter(s) for both
(2+1)- and (3+1)-dimensional EEs were presented in our unpublished
paper \cite{LTJH}.

Lie group theory is one of the most effective methods of seeking
exact and analytic solutions of physical systems. However, even for
a mathematician, it is still rather difficult to find a symmetry
group, especially non-Lie and non-local symmetry groups. So for
physicists, it would be more significant and meaningful to establish
a \em simple \rm method to obtain \em more general \rm symmetry
groups of nonlinear systems without using complicated group theory.

To our knowledge, there is little exact analytic understanding of
the vortices and circumfluence, although they are most general
observations in some physical fields; in particular, very rich
vortex structures exist in fluid systems. In fact, if one could find
the full symmetry groups of the EEs, then many kinds of exact vortex
and circumfluence solutions could be generated from some simple
trivial solutions.

This paper is an enlarged version of our earlier, unpublished paper
\cite{LTJH}. In section II, we first establish a simple direct
method to find a general group transformation theorem for the
(2+1)-dimensional EE, then utilize the theorem in some special cases
to obtain some solution theorems which lead to a quite general
symmetric vortex solution with some arbitrary functions. The
applications of the exact vortices and circumfluence solutions are
given in section III. It is indicated that the solutions can explain
the tropical cyclone (TC) eye, the track, and the relation between
the track and the background wind. The TC tracks can thus be
predicted by the relation. In section IV, beginning with a general
symmetry group theorem, the DT in \cite{Li} with zero spectral
parameter is extended to that with arbitrary spectral parameter. The
last section is a short summary and discussion.

\section{Space-time transformation group of the two-dimensional
EE.}

In the traditional theory, to find the Lie symmetry group of a given
nonlinear physical system, one has to first find its Lie symmetry
algebra and then use the Lie's first fundamental theorem to solve an
``initial" problem. If one utilizes the standard Lie group theory to
study the symmetry group of the two-dimensional EE, it is easy to
find that the only possible symmetry transformations are the
arbitrary time-dependent space and stream translations, constant
time translation, space rotation and scaling \cite{HF}.

Recently, for simplicity and finding \em more general \rm symmetry
groups, some types of new simple direct method without the use of
any group theory have been established for both Lax-integrable
\cite{group_CSF} and non-Lax-integrable \cite{group_JPA} models.

For the two-dimensional EE \eqref{e}, we have the following (weak)
Lax pair theorem.

{{\em Theorem 1   (Lax pair theorem \cite{Li}).} The
(2+1)-dimensional EE \eqref{e} possesses the weak Lax pair
\begin{eqnarray}
&& \omega_x\phi_{y}-\omega_y\phi_{x}=\lambda\phi,\label{Lax11}\\
&&\phi_t+\psi_x\phi_y-\psi_y\phi_x=0,  \label{Lax12}
\end{eqnarray}
with the spectral parameter $\lambda$.

{\em Proof.} To prove the theorem, we rewrite \eqref{Lax11} and
\eqref{Lax12} as
\begin{eqnarray}
&&L\phi=0,\qquad~ L\equiv [\omega,\ \cdot] -\lambda ,\label{L}\\
&&M\phi=0,\qquad M\equiv \partial_t+[\psi,\ \cdot]. \label{M}
\end{eqnarray}
It is straightforward that the compatibility condition of Eqs.
\eqref{L} and \eqref{M}, $LM-ML=0$, reads
\begin{eqnarray}
&&LM-M L = -[\omega_t,\ \cdot]-[\psi, \ [\omega,\
\cdot]]+[\omega,\ [\psi,\ \cdot]]=0. \label{ML}
\end{eqnarray}
Using the Jacobian identity for the commutator $[\cdot,\ \cdot]$
defined by \eqref{AB}
$$[A,\ [B,\ C]]+[B,\ [C,\ A]]+[C,\ [A,\ B]]=0,$$
Eq. \eqref{ML} becomes
\begin{eqnarray}
[\omega_t+[\psi,\ \omega],\ \cdot]=0.
 \label{cdot}
\end{eqnarray}
The theorem is proven. $\Box$

{\em Remark 1.} The theorem was proved in a slightly weak way in
\cite{Li1}, where the compatibility condition of Eqs. \eqref{Lax11}
and \eqref{Lax12} was
\begin{eqnarray}
[\omega_t+[\psi,\ \omega],\ \phi]=0,
 \label{cdot1}
\end{eqnarray}
with the requirement that $\phi$ was just the spectral function. However, in our new proof procedure, the
spectral function $\phi$ in Eq. \eqref{cdot1} can be replaced by any \em arbitrary \rm function.\\
{\em Remark 2.} In Theorem 1, the Lax pair is termed weak because
starting from the Lax pair, we can only prove Eq. \eqref{cdot}
instead of the EE \eqref{e} itself. For instance, Eq. \eqref{cdot}
is true for
$$\omega_t+[\psi,\ \omega]=c(t)$$
with $c(t)$ being an \em arbitrary \rm function of $t$. Therefore, all the conclusions obtained from the Lax
pair have to be treated carefully by substituting the final results to the original EE to rule out the
additional freedoms.

From Theorem 1, we know that the (2+1)-dimensional Euler equation is
weak Lax integrable. So we can apply the new direct method developed
in \cite{group_CSF} to find some complicated exact solutions from
some simple special trivial ones after ruling out the ambiguity
mentioned in remark 2.

Using the method in \cite{group_CSF}, we have the following
transformation theorem:

{\em Theorem 2.} {\em (Group Theorem).} If $\{\omega'(x,\ y,\ t),\
\psi'(x,\ y,\ t),\ \phi'(x,\ y,\ t)\}$ is a known solution of the
two-dimensional EE \eqref{e} and its Lax pair \eqref{Lax11} and
\eqref{Lax12} with the spectral parameter $\lambda'$, $\{\omega,\
\psi,\ \phi\}$ with
\begin{eqnarray}
\phi=\exp(g)\phi'(\xi,\ \eta,\ \tau)\equiv \exp(g)\phi'
\label{trans}
\end{eqnarray}
is a solution of Eq. \eqref{cdot} and its Lax pair with the spectral
parameter $\lambda$, if and only if the following three conditions
are satisfied:
\begin{eqnarray}
&&\left\{([\tau, \omega]\psi'_\xi+[\omega,
\eta])\lambda'-\lambda\omega'_\xi+[\omega,
g]\omega'_\xi\right\}\phi'\nonumber\\
 &&\qquad +\left\{[\tau,
\omega]\omega'_\tau+[\omega, \eta]\omega'_\eta+[\omega,
\xi]\omega'_\xi\right\}\phi'_\xi=0,\label{det1}\\
 &&\left\{\left[\eta_t+[\psi,  \eta]-(\tau_t+[\psi, \tau])
\psi'_\xi\right]\lambda'+(g_t+[\psi, g])\omega'_\xi\right\}\phi'\nonumber\\
&&\qquad+\left\{(\tau_t+[\psi, \tau])\omega'_\tau+(\eta_t+[\psi,
\eta])\omega'_\eta+(\xi_t+[\psi,
\xi])\omega'_\xi\right\}\phi'_\xi=0,\label{det2}\\
&&\omega=\psi_{xx}+\psi_{yy},\label{det}
\end{eqnarray}
where the arguments $\{x,\ y,\ t\}$ of the functions $\omega',\
\psi'$ and $\phi'$ have been transformed to $\{\xi,\ \eta,\
\tau\}$, and $\xi,\ \eta,\ \tau$ and $g$ are functions of $\{x,\
y,\ t\}$.}

{\em Proof.} Because $\{\omega'(x,\ y,\ t),\ \psi'(x,\ y,\ t),\
\phi'(x,\ y,\ t)\}$ is a solution of the EE and its Lax pair with
the spectral parameter $\lambda'$, then $\{\omega'(\xi,\ \eta,\
\tau),\ \psi'(\xi,\ \eta,\ \tau),\ \phi'(\xi,\ \eta,\ \tau)\}$
satisfies
\begin{eqnarray}
&&
\omega'_{\xi}\phi'_{\eta}-\omega'_{\eta}\phi'_{\xi}=\lambda'\phi',\label{Lax21}\\
&& \phi'_{\tau}+\psi'_{\xi}\phi'_{\eta}-\psi'_{\eta}\phi'_{\xi}=0,
\label{Lax22}
\end{eqnarray}
and
\begin{eqnarray}
\omega'_\tau+\psi'_\xi\omega'_\eta-\psi'_\eta\omega'_\xi=0.\label{O'}
\end{eqnarray}
Substituting Eq. \eqref{trans} into Eqs. \eqref{Lax11} and
\eqref{Lax12}, we have
\begin{eqnarray}
&&[\omega,\ \xi]\phi'_\xi +[\omega,\ \eta]\phi'_\eta +[\omega,\
\tau]\phi'_\tau +([\omega,\ g]-\lambda)\phi'=0,\label{e12a}
\\ &&
\big(\xi_t+[\psi,\ \xi]\big)\phi'_\xi +\big(\eta_t+[\psi,\
\eta]\big)\phi'_\eta +\big(\tau_t+[\psi,\ \tau])\phi'_\tau
+\big(g_t+[\psi,\ g]\big)\phi'=0.\label{e12}
\end{eqnarray}
Applying Eqs. \eqref{Lax21}, \eqref{Lax22} and \eqref{O'} to Eqs.
\eqref{e12a} and \eqref{e12} by ruling out the quantities
$\phi'_\tau$ and $\phi'_\eta$ yields Eqs. \eqref{det1} and
\eqref{det2}.

It is noted that Eq. \eqref{det} in Theorem 2 is only the definition
equation of the vorticity. Theorem 2 is proven. $\Box$

From Theorem 2, we have only three determinant equations for six
undetermined functions $\xi,\ \eta,\ \tau,\ \psi,\ \omega$ and $g$,
which means that the determinant equation system \eqref{det} is
underdetermined. Therefore, there exist abundant interesting exact
solutions. Here we consider two special interesting cases of Theorem
2.

{\em Corollary 1.} { If $\psi'(x,\ y,\ t)$ is a solution of the
 Poisson equation
\begin{eqnarray}
\omega_0=\psi'_{xx}+\psi'_{yy}
\end{eqnarray}
with a constant $\omega_0$, then $\{\omega,\ \psi\}$ is a solution
of \eqref{cdot} if the following three conditions hold:
\begin{eqnarray}
&&[\tau, \omega]\psi'_\xi+[\omega,
\eta]=0,\label{deta}\\
 &&\eta_t+[\psi,  \eta]-(\tau_t+[\psi, \tau])
\psi'_\xi=0,\label{detb}\\
&&\omega=\psi_{xx}+\psi_{yy},\label{detc}
\end{eqnarray}
where $\psi'\equiv \psi'(x,\ y,\ t)$ has been redefined as
$\psi'(\xi,\ \eta,\ \tau)$.}

{\em Proof.} It is clear that the EE \eqref{e} [and then Eq.
\eqref{cdot}] possesses a trivial constant vorticity solution
$\{\omega',\ \psi'\}=\{\omega_0,\ \psi'\}$ with $\psi'$ being a
solution of the Poisson equation. Substituting
$\omega'=\omega_0={\rm const}.$ into Theorem 2 results in the
Corollary 1 at once. $\Box$

{\em Corollary 2.} { If $\{\omega'(x,\ y,\ t),\ \psi'(x,\ y,\ t)\}$
is a known solution of the two-dimensional EE \eqref{e}, then
$\{\omega,\ \psi\}$ with the conditions
\begin{eqnarray}
&&[\tau, \omega]\omega'_\tau+[\omega, \eta]\omega'_\eta+[\omega,
\xi]\omega'_\xi=0,\label{Det1}\\
&&(\tau_t+[\psi, \tau])\omega'_\tau+(\eta_t+[\psi,
\eta])\omega'_\eta+(\xi_t+[\psi,
\xi])\omega'_\xi=0,\label{Det2}\\
&&\omega=\psi_{xx}+\psi_{yy}\label{Det}
\end{eqnarray}
is a solution of Eq. \eqref{cdot}, where the arguments $\{x,\ y,\
t\}$ of the functions $\omega'$ and $\psi'$ have been transformed
to $\{\xi,\ \eta,\ \tau\}$, and $\xi,\ \eta $ and $\tau$ are
functions of $\{x,\ y,\ t\}$.}

Corollary 2 can be readily obtained from Theorem 2 by taking
$\lambda'=\lambda=g=0$.

{\em Remark 3.} Corollary 1 and Corollary 2 are independent of the
Lax pair though they are derived by means of the Lax pair.

By solving Corollary 2, we can get the following theorem.

{\em Theorem 3 (Solution theorem).} The (2+1)-dimensional EE
possesses a special
 solution
$\{\omega,\ \psi\}$ with
\begin{eqnarray}
&&\omega= F(f(x,\ y,\ t))\equiv F,\label{oF}\\
&&\psi = G(f(x,\ y,\ t),\ t)-\int^x\frac{f_t(z,\ r,\ t)}{f_r(z,\ r,\ t)}{\rm d}z \equiv G+h,\label{Gh}
\end{eqnarray}
 where
$f\equiv f(x,\ y,\ t), \ h\equiv h(x,\ y,\ t), \ F(f)$ and $G(f,\
t)$ are functions of the indicated variables, the variable
$r=r(x,\ y,\ z,\ t)$ is determined by
$$f(z,\ r,\ t)=f(x,\ y,\ t)$$
 and the
functions $F,\ G$ and $f$ ($h$) are linked by the following
constrained condition
\begin{eqnarray}
F=G_{ff}(f_x^2+f_y^2)+G_f(f_{xx}+f_{yy})+h_{xx}+h_{yy}.\label{FG}
\end{eqnarray}

{\em Proof.} After rewriting Eqs. \eqref{Det1} and \eqref{Det2} as
$$ [\omega,\ \omega']=0,\eqno(26')$$
$$ \omega'_t+[\psi,\ \omega']=0,\eqno(27')$$
it is not difficult to find that the general solution of Eq.
\eqref{Det1} [i.e. Eq. (26')] is
\begin{eqnarray}
\omega=F(\omega',\ t). \label{oF1}
\end{eqnarray}
Though $\omega'(x,\ y,\ t)$ should be an exact known solution of the
EE, $\omega'(\xi,\ \eta,\ \tau)\equiv f$ can still be considered as
an arbitrary function of $\{x,\ y,\ t\}$ due to the fact that $\xi,\
\eta$ and $\tau$ are all undetermined arbitrary functions of $\{x,\
y,\ t\}$. Then Eq. \eqref{oF1} becomes
\begin{eqnarray}
\omega=F(f,\ t). \label{oF1a}
\end{eqnarray}
The general solution of Eq. (27') [or Eq. \eqref{Det2}] is rightly
Eq. \eqref{Gh}, while Eq. \eqref{FG} is just the direct substitution
of Eqs. \eqref{oF1a} and \eqref{Gh} to Eq. \eqref{Det}.

Finally, to rule out the ambiguity brought by the weak Lax pair by
substituting Eqs. \eqref{oF1a} and \eqref{Gh} into Eq. \eqref{e},
one can find that Eq. \eqref{oF1a} with Eq. \eqref{Gh} is really a
solution of the EE \eqref{e} only if $F(f,t)=F(f)$. Theorem 3 is
proven. $\Box$

Because of the arbitrary function $f$, we can obtain many physically interesting solutions from Theorem 3. For
instance, if the arbitrary function $f$ is assumed to be the form
\begin{eqnarray}
f=(x-x_0)^2+(y-y_0)^2+h_0\equiv r+h_0 ,\label{f}
\end{eqnarray}
where $x_0,\ y_0$ and $h_0$ are all arbitrary functions of $t$, then
we readily have the following special solution theorem.

{\em Theorem 4 (Special solution theorem).} The (2+1)-dimensional EE
\eqref{e} possesses an exact solution
\begin{eqnarray}
\psi&=&y_{0t}x-x_{0t}y+F_1\ln r+F_2-\frac{1}2h_{0t}\tan^{-1}\frac{x-x_0}{y-y_0} +\frac14\int
\frac{F(r+h_0)}r{\rm d}r,\label{vortex}\\
\omega&=&F_r(r+h_0),\label{vortex11}
\end{eqnarray}
where
 $x_0,\ y_0,\ h_0,\ F_1$ and $F_2$ are arbitrary functions of
$t$, and $F\equiv F(r+h_0)$ is an arbitrary function of $r+h_0$.

The intrusion of many arbitrary functions into the exact solution
\eqref{vortex} allows us to find various vortex and circumfluence
structures by selecting them in different ways.

In the solution \eqref{vortex}, the first two terms
$$y_{0t}x-x_{0t}y$$
represent the background wind (induced flow) with the time-dependent
velocity field
$$\vec{u}=\{x_{0t},\ y_{0t}\}.$$

The third term ($F_1$-dependent)
\begin{eqnarray}
F_1\ln r, \label{F_1}
\end{eqnarray}
corresponds to a time-dependent singular vortex. The detailed
velocity field with
\begin{eqnarray}
F_1=1,\qquad x_0=y_0=0 \label{F11}
\end{eqnarray}
is shown in Fig. 1. All the quantities used in the figures of this
paper are dimensionless except for the special indication in Fig.
8.

\input epsf
\begin{figure}
\epsfxsize=7cm\epsfysize=7cm\epsfbox{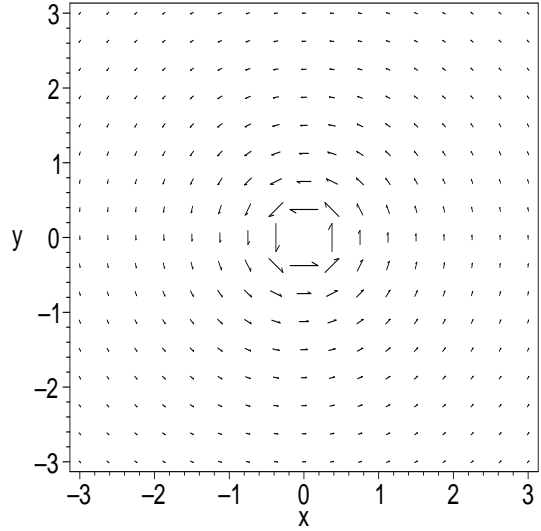}
 \caption{The structure of the singular vortex expressed by \eqref{F_1} with the parameters \eqref{F11}.
 The length of the arrow stands for the strength of
  the velocity field and the values from inside to outside are 16/3, 8/3, 16/9, 4/3,
 16/15, 8/9, 16/21 and 2/3, respectively.}
\end{figure}

The fourth term $F_2$ is trivial because of the existence of the
time-dependent translation freedom when one introduces the potential
of the velocity---i.e., the stream function.

The fifth term ($h_{0t}$-dependent)
\begin{eqnarray}
\frac{1}2h_{0t}\tan^{-1}\frac{x-x_0}{y-y_0} \label{hole}
\end{eqnarray}
is related to a hole [Fig. 2(a)] or a source [Fig. 2(b)].

\input epsf
\begin{figure}
\epsfxsize=7cm\epsfysize=7cm\epsfbox{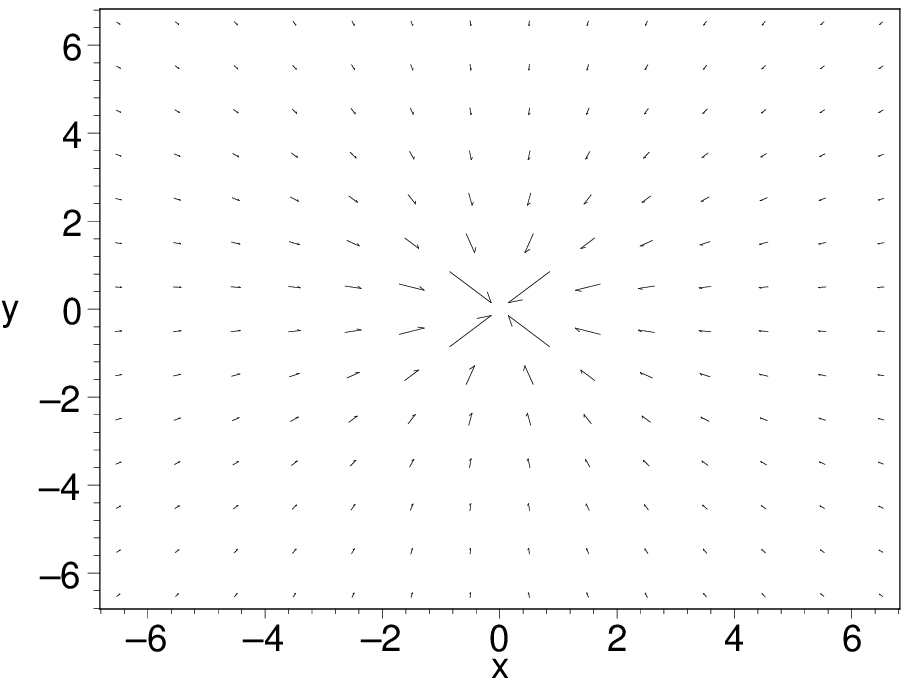}
\epsfxsize=7cm\epsfysize=7cm\epsfbox{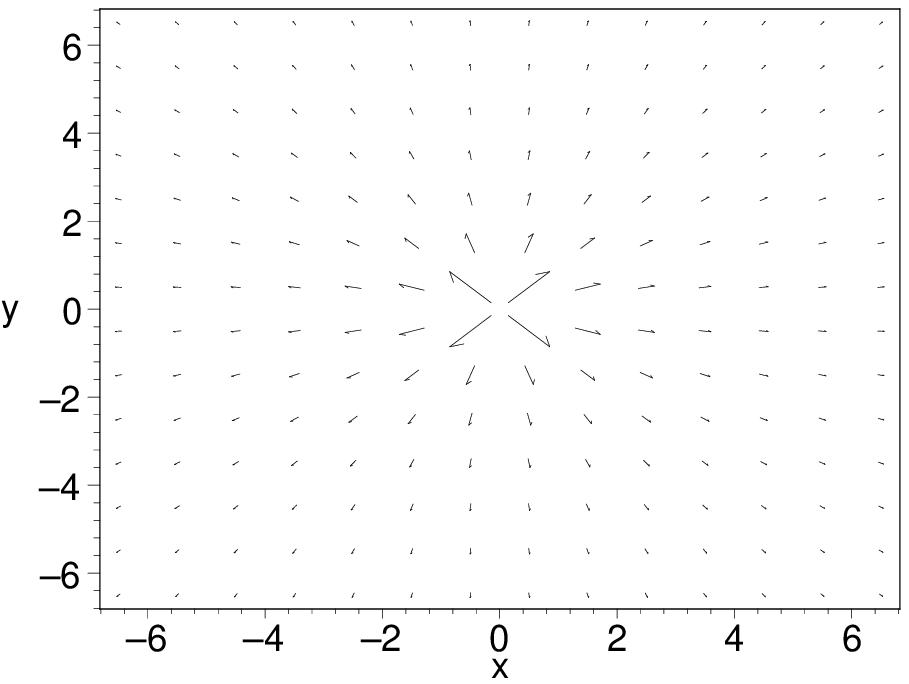}
 \caption{(a). The structure of the hole expressed by \eqref{hole} with $x_0=y_0=0,h_0=t$.
 (b). The source structure of \eqref{hole} with $x_0=y_0=0$ and $h_0=-t$. The length of the arrows
 expresses the strength of
 the velocity field and the values from inside to outside are 0.71, 0.32, 0.20, 0.14, 0.11, 0.09, and
 0.08,
 respectively, both for (a) and (b).}
\end{figure}

The last term of Eq. \eqref{vortex}
\begin{eqnarray}
\frac14\int \frac{F(r+h_0)}r{\rm d}r,\  \label{Vot}
\end{eqnarray}
is the most interesting because it is related to abundant vortex
structures due to the arbitrariness of the function $F$. Here are
some special examples based on the different selections of the
arbitrary function.

(i) {\em  Lump-type vortices.} If the function $F(r)$ is a rational
solution of $r$,
\begin{eqnarray}
F(r)=\frac{\sum_{i=0}^Na_ir^i}{\sum_{i=0}^Nb_ir^i}\equiv
\frac{P(r)}{Q(r)} \label{rational}
\end{eqnarray}
with the conditions $b_N\neq0$ and $Q(r)\neq 0$ for all $r\geq 0$,
then the solution \eqref{Vot} becomes an analytical lump-type vortex
and/or circumfluence solution for the velocity field. Figure 3
displays a special lump-type vortex structure of the velocity field
described by \eqref{Vot} with
\begin{eqnarray}
F(r)= \frac{10 r}{1+10r^2}, \ x_0=y_0=0. \label{rational1}
\end{eqnarray}

\input epsf
\begin{figure}
\epsfxsize=7cm\epsfysize=7cm\epsfbox{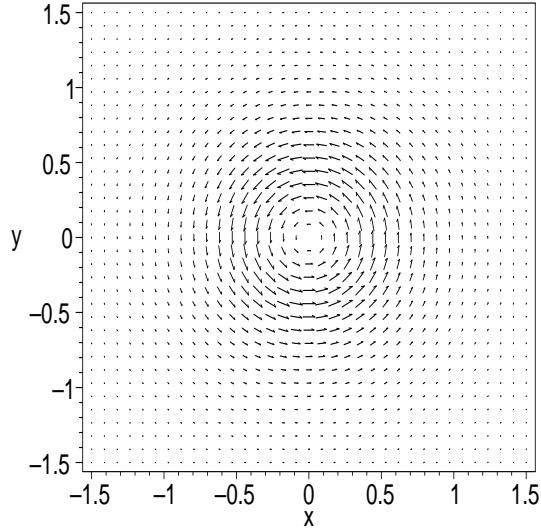}
 \caption{A typical lump-type vortex expressed by \eqref{Vot} with \eqref{rational1}. The strength of
 the velocity field is expressed by the length of the arrows and the values from inside to outside
 are 0.44, 0.87, 1.26, 1.52, 1.60, 1.48, 1.26, 1.01, 0.80, 0.62, 0.49, 0.39, 0.31, 0.25, 0.21, 0.17, and
 0.14,
 respectively.}
\end{figure}

(ii) {\em Dromion-type vortices.} When the function $F(r)$ is fixed
as a rational function of $r$ multiplied by an exponentially
decaying factor---for instance,
\begin{eqnarray}
F(r)=\frac{\sum_{i=0}^Na_ir^i}{\sum_{i=0}^Nb_ir^i}\exp(-c^2r),
\label{exp}
\end{eqnarray}
with arbitrary constants $a_i,\ b_i$ and $c$---then \eqref{Vot}
turns into an analytical dromion-type vortex and/or circumfluence
solution. Figure 4 exhibits a particular dromion-type vortex
structure of Eq. (40) with
\begin{eqnarray}
F(r)=r\exp(-r), \ x_0=y_0=0. \label{exp1}
\end{eqnarray}
\input epsf
\begin{figure}
\epsfxsize=7cm\epsfysize=7cm\epsfbox{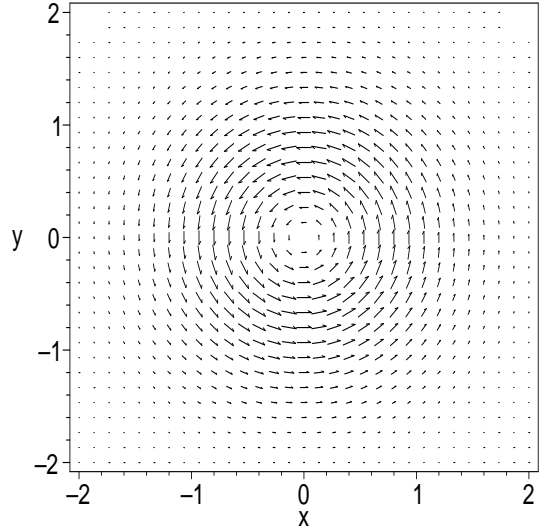}
 \caption{A typical dromion-type vortex expressed by \eqref{Vot} with Eqs. \eqref{exp1}. The strength of
 the velocity field is expressed by the length of the arrows and the values from inside to outside
 are 0.06, 0.12, 0.17, 0.20, 0.21, 0.20, 0.19, 0.17, 0.14, 0.11, 0.08, 0.06, 0.04, 0.032, and 0.02, respectively.}
\end{figure}

(iii) {\em Ring solitons and circumfluence.} Recently, some kinds of
ring soliton solutions were discovered \cite{ring,ring1}. It is
interesting that the basin and plateau types of ring solitons may be
responsible for the circumfluence solution for fluid systems
described by the EE. For instance, if  $F(r)$ is assumed to have the
property
$$\left.
\frac{{\rm d}^i F(r)}{{\rm d} r^i}\right|_{r=0}=0,\qquad i=0,\ 1,\
...,\ n,$$ for $n\geq 2$, then \eqref{Vot} expresses the
circumfluence for the velocity field and the basin- or plateau-type
ring
soliton for the stream function. Figure 5(a) exhibits a special picture with 
\begin{eqnarray}
F(r)=-4{r^{2}e^{-r}} \label{circulation}
\end{eqnarray}
of the circumfluence structure for the velocity field, Fig. 5(b)
displays the corresponding basin-type ring soliton shape for the
stream function $\psi$, and Fig. 5(c) shows the structure of the
vorticity.

\input epsf
\begin{figure}
\epsfxsize=7cm\epsfysize=7cm\epsfbox{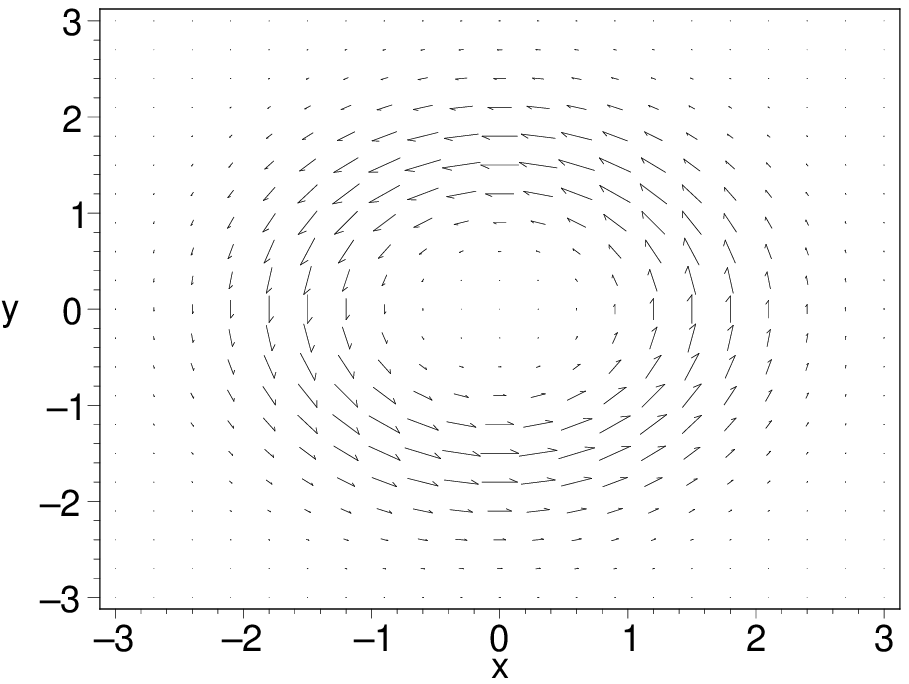}
\epsfxsize=7cm\epsfysize=7cm\epsfbox{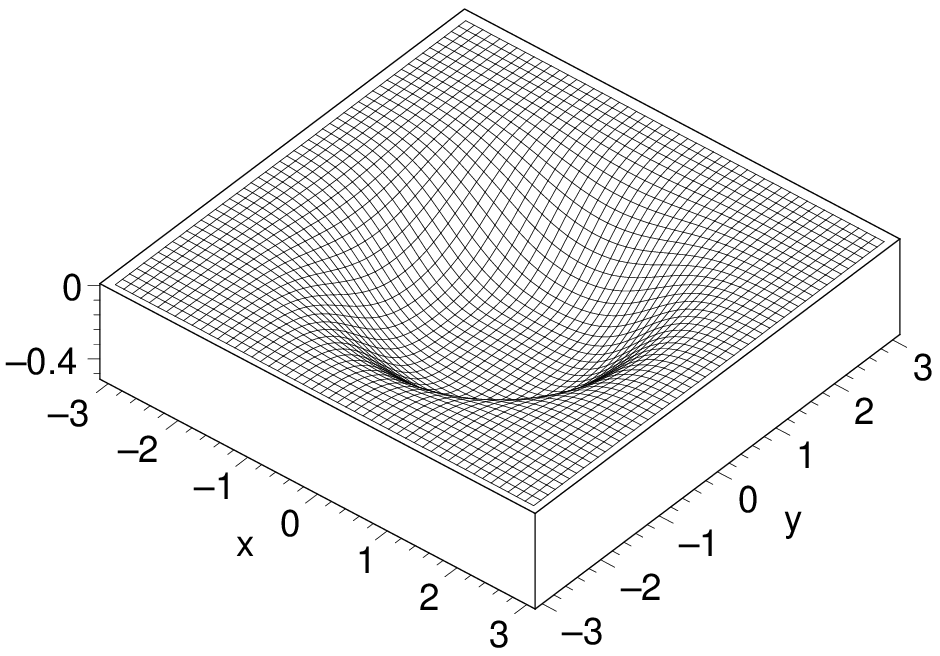}
\epsfxsize=7cm\epsfysize=7cm\epsfbox{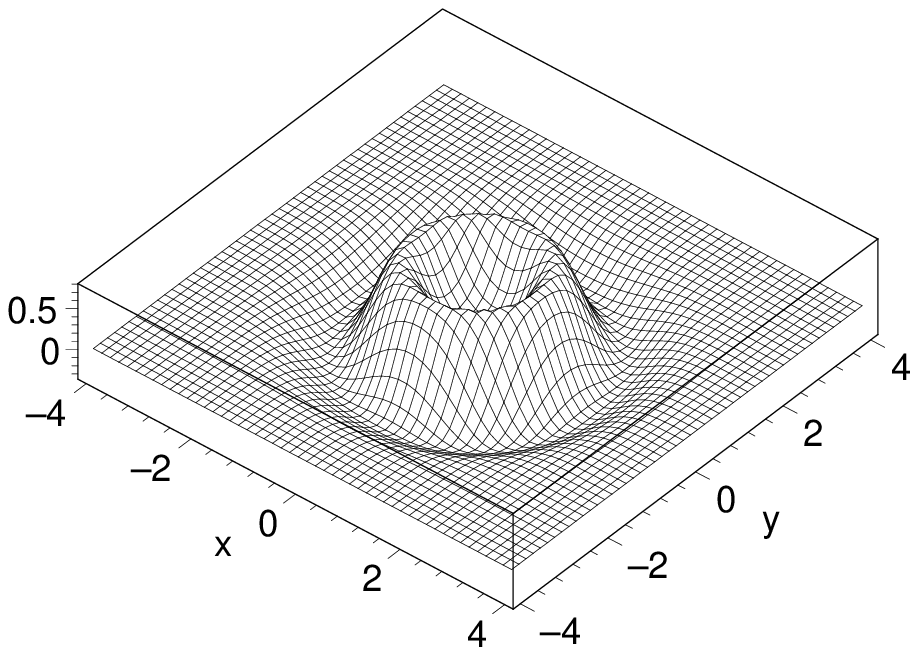}
 \caption{(a)
A field plot of the circumfluence \eqref{Vot} with Eq.
\eqref{circulation} for the velocity $\{u,\ v\}$. The strength of
 the velocity field is expressed by the length of the arrows, and the values from inside to outside
 are 0.001, 0.027, 0.131, 0.295, 0.400, 0.370, 0.248, 0.125, 0.049, and 0.015, respectively.
 (b) The
corresponding basin-type ring soliton for the stream function
$\psi$ related to (a). (c) The corresponding bow-type ring soliton
for the vorticity $\omega$. }
\end{figure}

\section{Applications to Hurricane Katrina 2005.}
It is demonstrated in the last section that the exact solutions
\eqref{vortex}-\eqref{vortex11} have quite rich structures. Due to
the richness of the solution structures and wide applications of the
vortex in various fields such as fluids, plasma, oceanic and
atmospheric dynamics, cosmography, astrophysics, condensed matter,
etc. \cite{fluid}--\cite{H}, our results may be applied in all these
fields. For instance, in oceanic and atmospheric dynamics, the
analytical solution \eqref{vortex} can be used to approximately
describe TCs which possess increasing destructiveness over the past
30 years \cite{typhoon}. The relatively tranquil part, the center of
the circumfluence shown in Fig. 5(a) is responsible for the TC eye
\cite{eye}.

 To describe different types of vortexes, one may select different types of
function $F(r)$. To qualitatively and even quantitatively
characterize TCs, we may require that
 $F(r)$ have the form
\begin{eqnarray}
F(r)=\pm a^2{r^{1+b^2}e^{-c^2\sqrt{r}}}, \label{Hurricane}
\end{eqnarray}
with constants $a,\ b$ and $c$. In Eq. \eqref{Hurricane}, the signs
``$+$" and ``$-$" dictate the TCs of the northern and southern
hemisphere, respectively. The constants $a,\ b$ and $c$ are
responsible for the strength, the size of the TC's eye, and the
width of the TC.

The corresponding stream function related to the selection
\eqref{Hurricane} reads
\begin{eqnarray}
\psi&=&y_{0t}x-x_{0t}y+\frac{a^2(1+b^2)}{2c^{4+6b^2}\sqrt{\pi}}r^{-\frac{b^2}2}e^{-c^2\sqrt{r}}
\left\{(2b^2+1)\left[4^{b^2}b^2\Gamma(b^2)\Gamma\left(b^2+\frac12\right)\right.\right.\nonumber\\
&&\left.\left.-\sqrt{\pi}\Gamma(2b^2+1,c^2\sqrt{r})\right]e^{c^2\sqrt{r}}-r^{b^2+\frac12}c^{4b^2+2}\right\},\label{psi}
\end{eqnarray}
where $\Gamma(z)$ and $\Gamma(a,z)$ are the usual Gamma and
incomplete Gamma functions, respectively.

For more concreteness, we take Hurricane Katrina 2005 as an
illustration.

Figure 6 is the satellite image downloaded from the web
http://www.katrina.noaa.gov/ satellite/satellite.html \cite{noaa}
for Hurricane Katrina 2005 at 14:15, August 29, 2005, Coordinated
Universal Time (UTC).

\input epsf
\begin{figure}
\epsfxsize=7cm\epsfysize=4cm\epsfbox{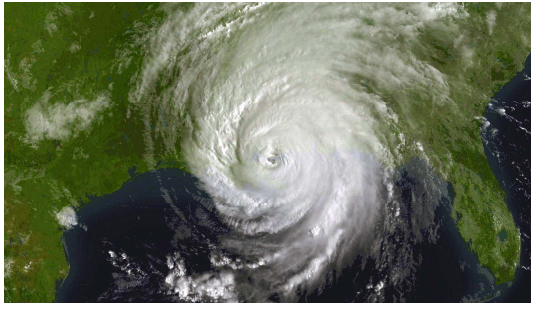}
 \caption{(Color online) \small
The satellite image of Hurricane Katrina 2005 at 14:15, August 29,
2005, Coordinated Universal Time.}
\end{figure}

To fix the constants $a,\ b$ and $c$ in Eq. \eqref{Hurricane} for
Hurricane Katrina 2005 shown in Fig. 6, we need know the strength
(the maximum wind speed), the eye size, and the width of TC Katrina
at 14:15, August 29, 2005 UTC. The strength of Katrina can be found
in several websites. The data of Table I are downloaded from
\cite{G}.

\newpage
Table I. Data of Hurricane Katrina downloaded from \cite{G}.

\begin{tabular}{ l c  c  c  c  c  c  c}
  \hline
   \hline
Time (UTC)     &  W. Long&  N. Lat.  &  MPH  &  Time (UTC)     & W.
Long&  N. Lat.  & MPH \\ \hline
 2005 Aug 23 21:00  & 75.50  & 23.20   &  35  & 2005 Aug 27 03:00 &  83.60 &  24.60 & 105   \\
   2005 Aug 24 00:00  & 75.80 &  23.30 &  35  & 2005 Aug 27 06:00 &  84.00 &  24.40 & 110  \\
   2005 Aug 24 03:00  & 76.00 &  23.40 &  35  &2005 Aug 27 09:00 &  84.40 &  24.40 & 115   \\
   2005 Aug 24 06:00  & 76.00 &  23.60 &  35  &   2005 Aug 27 12:00 &  84.60 &  24.40 & 115   \\
   2005 Aug 24 09:00  & 76.40 &  24.00 &  35  &2005 Aug 27 15:00 &  85.00 &  24.50 & 115   \\
   2005 Aug 24 12:00  & 76.60 &  24.40 &  35  &2005 Aug 27 18:00 &  85.40 &  24.50 & 115   \\
   2005 Aug 24 15:00  & 76.70 &  24.70 &  40   &2005 Aug 27 21:00 &  85.60 &  24.60 & 115   \\
   2005 Aug 24 18:00  & 77.00 &  25.20 &  45  &2005 Aug 28 00:00 &  85.90 &  24.80 & 115   \\
   2005 Aug 24 21:00  & 77.20 &  25.60 &  45  &2005 Aug 28 03:00 &  86.20 &  25.00 & 115  \\
   2005 Aug 25 00:00  & 77.60 &  26.00 &  45  &2005 Aug 28 06:00 &  86.80 &  25.10 & 145   \\
   2005 Aug 25 03:00  & 78.00 &  26.00 &  50  &2005 Aug 28 09:00 &  87.40 &  25.40 & 145   \\
   2005 Aug 25 06:00  & 78.40 &  26.10 &  50  & 2005 Aug 28 12:00 &  87.70 &  25.70 & 160   \\
   2005 Aug 25 09:00  & 78.70 &  26.20 &  50  & 2005 Aug 28 15:00 &  88.10 &  26.00 & 175   \\
   2005 Aug 25 12:00  & 79.00 &  26.20 &  50  &2005 Aug 28 18:00 &  88.60 &  26.50 & 175  \\
   2005 Aug 25 15:00  & 79.30 &  26.20 &  60  & 2005 Aug 28 21:00 &  89.00 &  26.90 & 165   \\
   2005 Aug 25 17:00  & 79.50 &  26.20 &  65   &2005 Aug 29 00:00 &  89.10 &  27.20 & 160   \\
   2005 Aug 25 19:00  & 79.60 &  26.20 &  70   &2005 Aug 29 03:00 &  89.40 &  27.60 & 160   \\
   2005 Aug 25 21:00  & 79.90 &  26.10 &  75   &2005 Aug 29 05:00 &  89.50 &  27.90 & 160   \\
   2005 Aug 25 23:00  & 80.10 &  25.90 &  80  &2005 Aug 29 07:00 &  89.60 &  28.20 & 155   \\
   2005 Aug 26 01:00  & 80.40 &  25.80 &  80  & 2005 Aug 29 09:00 &  89.60 &  28.80 & 150   \\
   2005 Aug 26 03:00  & 80.70 &  25.50 &  75  & 2005 Aug 29 11:00 &  89.60 &  29.10 & 145   \\
   2005 Aug 26 05:00  & 81.10 &  25.40 &  70 &  2005 Aug 29 13:00 &  89.60 &  29.70 & 135   \\
   2005 Aug 26 07:00  & 81.30 &  25.30 &  70   & 2005 Aug 29 15:00 &  89.60 &  30.20 & 125    \\
   2005 Aug 26 09:00  & 81.50 &  25.30 &  75   &2005 Aug 29 17:00 &  89.60 &  30.80 & 105    \\
   2005 Aug 26 11:00  & 81.80 &  25.30 &  75  & 2005 Aug 29 19:00 &  89.60 &  31.40 &  95   \\
   2005 Aug 26 13:00  & 82.00 &  25.20 &  75  &2005 Aug 29 21:00 &  89.60 &  31.90 &  75   \\
   2005 Aug 26 15:00 &  82.20 &  25.10 &  80   &2005 Aug 30 00:00 &  88.90 &  32.90 &  65   \\
   2005 Aug 26 15:30  & 82.20 &  25.10 & 100   & 2005 Aug 30 03:00 &  88.50 &  33.50 &  60   \\
   2005 Aug 26 18:00  & 82.60 &  24.90 & 100   & 2005 Aug 30 09:00 &  88.40 &  34.70 &  50   \\
   2005 Aug 26 21:00 &  82.90 &  24.80 & 100   & 2005 Aug 30 15:00 &  87.50 &  36.30 &  35    \\
  \hline
\end{tabular}

From Table I, we know that the maximum wind speed of the Katrina
2005 at 14:15, August 29, 2005 UTC is about 130 mph (miles per
hour)---i.e.,
\begin{eqnarray}
v_{\max}\sim 130 \mbox{\rm\ mph} \sim 200 \mbox{\rm\ km \ ph} \sim
2 \mbox{\rm\ degree\ ph}, \qquad v\equiv \sqrt{\psi_y^2+\psi_x^2}.
\label{strength}
\end{eqnarray}

Comparing Katrina's satellite image shown in Fig. 6 with the map of
New Orleans---say the map shown in Fig. 7 downloaded from
\cite{track}---one can estimate that the eye size ($E$) is about
$$E \sim 1  \mbox{\ \rm degree}  \sim 100  \mbox{\ \rm km}$$
and the width ($W$) of the hurricane is approximately
$$W\approx 10  \mbox{\ \rm degree} \approx 1000 \mbox{\ \rm km}$$
for Katrina at 14:15, August 29, 2005 UTC. Using these data, we can
find that the stream function of Katrina 2005 near the time at
14:15, August 29, 2005 UTC can be approximately described by
\begin{eqnarray}
\psi_{\mbox{Katrina}}\approx y_{0t}x-x_{0t}y -4
(2+2\sqrt{r}+r)\exp(-\sqrt{r}),\ \quad (r\equiv
(x-x_0)^2+(y-y_0)^2),\label{Katrina}
\end{eqnarray}
which corresponds to the parameter selections
$$a^2 \sim 8,\ b^2\sim  0.5,\ c^2 \sim
1, $$ in Eq. \eqref{Hurricane}. In the real case, the quantities
$a,\ b$ and $c$ should be time dependent. So the description here is
only approximate because it is only a solution of the EE instead of
the NS equation.

 If the strength $v_{\max}$, the size of the hurricane eye $E$, and the width $W$
are assumed to have some errors,
\begin{eqnarray}
v_{\max} \approx 130 \pm 10\ \mbox{\rm mph},\ E \approx 1 \pm 0.2
\mbox{\ \rm degree},\ W\approx 10 \pm 2 \mbox{\ \rm
degree},\label{error}
\end{eqnarray}
then the parameters $a,\ b$ and $c$ in Eq. \eqref{Hurricane} or
\eqref{psi} have the ranges
\begin{eqnarray}
 a^2 \approx 6.75 \sim 9.35,\ b^2\approx  0.25 \sim 1.25,\ c^2
\approx 0.9 \sim 1.2.\label{error1}
\end{eqnarray}

\input epsf
\begin{figure}
\epsfxsize=7cm\epsfysize=5.5cm\epsfbox{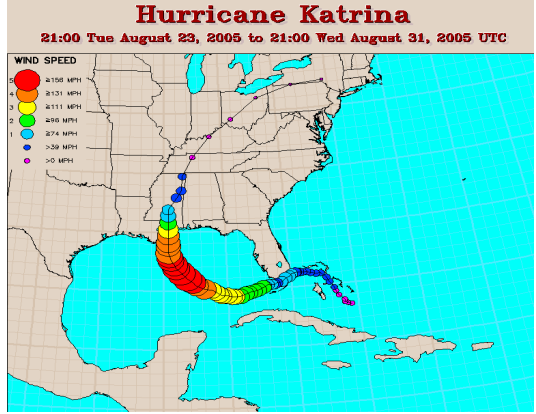}
 \caption{\small (Color online)
The map with the longitude and latitude degree coordinates near the
New Orleans and the real track of Hurricane Katrina 2005 from 21:00
Tues., August 23, 2005, to 21:00 Wednes., August 31, 2005 UTC.}
\end{figure}

In \eqref{Katrina}, $x_0\equiv x_0(t)$ and $y_0\equiv y_0(t)$ can be
obtained from the data in Table I. According to Table I, we can find
the theoretical fit of hurricane Katrina 2005 from 22:00, August 25,
2005 to 15:00, August 30, 2005 can be approximately described by
\begin{eqnarray}
x_0=0.00022t^2-0.14t-79,\ y_0=0.00073t^2-0.07t+26 \label{x0y01}
\end{eqnarray}
before 21:00, August 27, 2005 and
\begin{eqnarray}
x_0=0.0029t^2-0.5t-68,\ y_0=0.0023t^2-0.2t+29 \label{x0y02}
\end{eqnarray}
 after
21:00, August 27, 2005. In Eqs. \eqref{x0y01} and \eqref{x0y02}, the
units of $x_0,\ y_0$ and $t$ are longitude degree, latitude degree,
and hour, respectively, while the initial time $t=0$ is taken as
22:00, August 25, 2005 UTC.

{\em Remark 4.} If we fit the track only for the [Longitude,
Latitude] positions, we may get a better fit without using any
switch point. However, if we fit the track not only for the
positions but also for times, we have to select some switch points.
Physically speaking, when we use a parabolic line such as
\eqref{x0y01} to fit the track of a TC, we have to assume that the
TC moves under a constant force during the fit time period. The
necessary selections of the switch points are caused by the fact
that the driven force of the TC is time dependent. Here we find that
if we select 21:00, August 27, 2005 as a switch point, then Eqs.
\eqref{x0y01} and \eqref{x0y02} can fit the track quit well (the
square error [see later, Eq. \eqref{Delta}] becomes smallest). This
means that the TC is approximately driven by two constant forces
before and after the turning time, respectively. Actually,
\emph{approximately} speaking, after this switch point, the TC
becomes stronger and stronger (see Fig. 7 and/or Table I).

Figure 8 describes the velocity field of Katrina at 14:15, August
29, 2005 UTC ($t=88.15$) when the stream function is given by Eqs.
\eqref{Katrina} with \eqref{x0y02}.

\input epsf
\begin{figure}
\epsfxsize=7cm\epsfysize=7cm\epsfbox{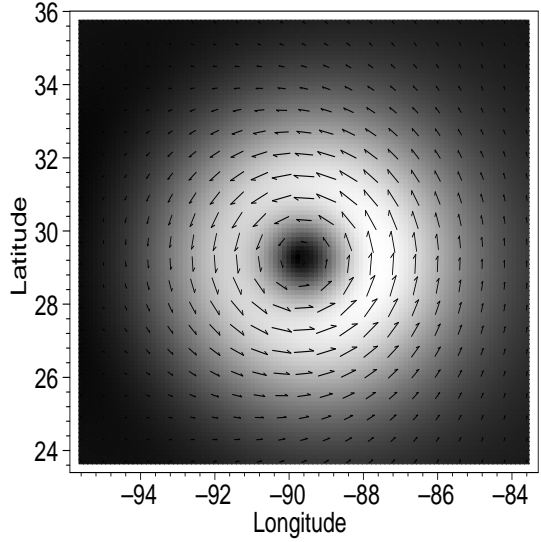}
 \caption{\small
 The field and density plot for the velocity field Hurricane
Katrina 2005 at 14:15/29/08 described by Eqs. \eqref{Katrina} with
\eqref{x0y02}. The strength of
 the velocity field is expressed by the length of the arrows and the values from left to
 right at $y=0$ are 0.14, 0.30, 0.53, 0.82, 1.17, 1.55, 1.86, 1.95, 1.60, 0.69, 0.18,
 0.89, 1.87, 2.33, 2.31, 2.03, 1.66, 1.30, 0.99, 0.75, and 0.58 (degree per hour), respectively.}
\end{figure}

In addition, the solution \eqref{vortex}-\eqref{vortex11} also
provides a relation between the TC track given by $\{x_0,\ y_0\}$
and the strength of the background wind (steering flow).

The stream function of the steering flow, $\psi_s$, can be obtained
by eliminating the TC term (vortex term) in Eq. \eqref{vortex} with
$F_1=F_2=h_0=0$ by setting $F=0$ and then the velocity field flow
$\overrightarrow{u}_s$ of the background wind reads
\begin{eqnarray}
\overrightarrow{u}_s=\{-\psi_{sy},\ \psi_{sx}\}= \{x_{0 t},\ y_{0
t}\}. \label{v}
\end{eqnarray}

This fact implies that once the background wind, or the
large-scale steering flow in the upper air, is known then the
motion of the hurricane center can be obtained.

Inversely, if the motion of the hurricane center is known then the
steering flow will be obtained at the same time. Therefore,
 if the position \{$x_0$, $y_0$\} of the hurricane
center is determined, in a not very long time (say, shorter than
one day), one can
 consider that the velocity of the TC will approximately
keep the latest known velocity and then the TC's new position
$\{x_1(t_1),\ y_1(t_1)\}$ at time $t_1$ can be determined by using
\begin{eqnarray}
x_1(t_1)=x_0(t_0)+x_{0t}(t_0)(t_1-t_0),\quad
y_1(t_1)=y_0(t_0)+y_{0t}(t_0)(t_1-t_0). \label{xy}
\end{eqnarray}

The concrete steps to predict the track and position of a TC are as
follows.

(i) Get the original known position data of a TC from professional
meteorologic web site. The concrete position data of a happening TC,
say, Katrina, can be read off from some web sites, say, \cite{G,G1},
which are given by some international satellites and updated every
six hours and usually three hours (or every hour) close to the
landing time.

(ii) Take the coordinate of the fit track. From the web site we can
get the position described by longitude and latitude. Because the
TCs happen in a quite small area compared to the whole Earth, the
fit curve can be taken in a two-dimensional plane. To simplify, the
longitude and latitude are defined as $X$ axis $Y$ axes,
respectively. The first time recorded on the web site is set as
initial time, and the following times are added in order by the time
interval.

(iii) Fit the function curve and forecast the track and position.
From the first few known positions, it is easy to calculate the fit
curve which is the function of time $t$. Usually we can take it
possesses the polynomial forms of the time $t$, say,
$\{X=x_0(t)=\sum_{i=0}^{N}a_it^i,\ Y=y_0(t)=\sum_{i=0}^{N}b_it^i\}$.
In this paper, we take $N=2$. Finally we should minimize the square
error, $\Delta$, among the fit track and the real track
\begin{eqnarray}
\Delta\equiv
\sum_{j=n_1}^{n_2}\left[(x_0(t_{j})-x_{j})^2+(y_0(t_j)-y_j)^2\right]\label{Delta}
\end{eqnarray}
by fixing the constants $a_i$ and $b_i$, where $\{x_j,\ y_j\}$ and
$\{x_0(t_j),\ y_0(t_j)\}$ are the real and fit positions of the
hurricane center at time $t_j$, $n_1,\ n_1+1,\ ...,\ n_2$ are
related to the points used to fit the theoretical track $\{x_0(t),\
y_0(t)\}$, $t=t_{n_2}$ corresponds to the time to make the further
prediction. Usually, we take $t_{n_2}-t_{n_1}\sim 24\ ({\mbox{\rm
hours}})$ that means the earlier history can be neglected to the
hurricane track.

Based on the above descriptions, we can use first few known position
data of the hurricane center to predict the possible position of the
hurricane some hours later.

Figure 9 displays an example on the TC track [Fig. 9(a)] and the
related background wind field [Figs. 9(b)--9(f)]. The zigzag line in
Fig. 9(a) is the real track of Katrina 2005 from 21:00/25/08 to
15:00/30/08 (data are read off from \cite{G}), and the solid line is
our fit (by using all the data in Table I) given by Eqs.
\eqref{x0y01} and \eqref{x0y02}. Figures. 9(b)--9(f) reveal the
corresponding steering flows at five different times. It is shown
that the background wind leads to the change of the direction of the
TC track. According to the relation between the track and the
steering flow, we can predict the TC track by using several
beginning data. The cross points in Fig. 9(a) are our predicted
track 6 h before the real one. The same idea has been applied to
typhoon Chanchu 2006 \cite{JiaLou} and Hurricane Andrew 1992
\cite{LTJH}.

\input epsf
\begin{figure}
\epsfxsize=14cm\epsfysize=7cm\epsfbox{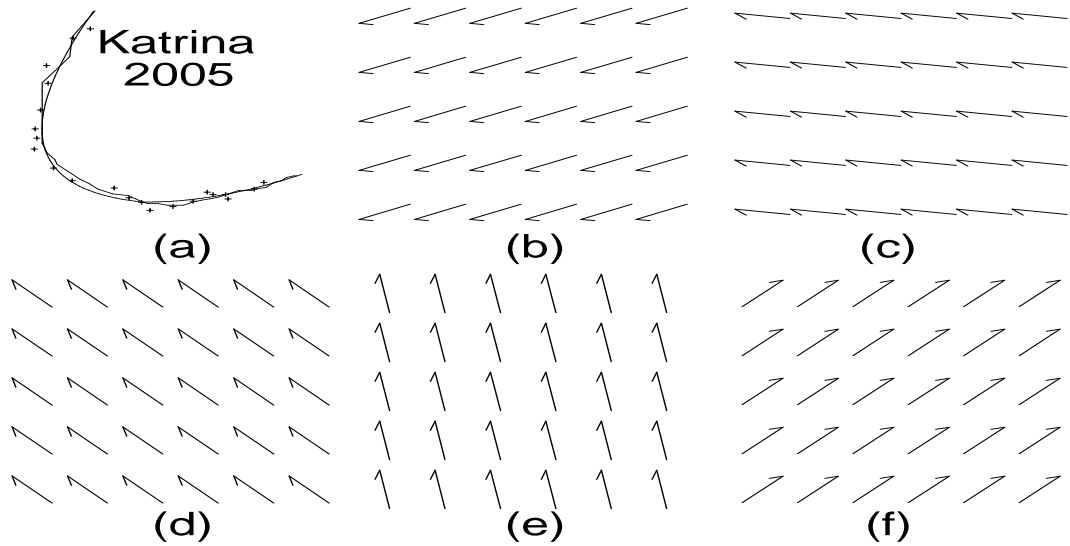} \caption{\small
(a). The zigzag line stands for the real TC track of Hurricane
Katrina 2005 from 21:00/25/08 to 15:00/30/08 for time and from
[79.9,26.2] to [87.5,36.3] for [Longitude, Latitude] position, the
smooth line is the fit track by using all the data in Table I and
the cross points express our predicted track six hours before the
real time. (b)--(f). The corresponding steering flows at
23:00/25/08, 21:00/27/08, 15:00/28/08, 05:00/29/08 and 00:00/30/08
with the strengths 1.33, 1.92, 2.92, 2.67, and 1.08 (degree per
hour), respectively.}
\end{figure}

\section{Darboux transformation of the (2+1)-dimensional EE}

In Sec. II, we have established a general group theorem (Theorem 2)
for the (2+1)-dimensional EE, which can yield various solutions.
Furthermore, two solution corollaries on the (2+1)-dimensional EE
are obtained by utilizing particular seed solutions.

It is noted that the weak DT theorems given in \cite{Li,LTJH,LouLi}
are special cases of the general group Theorem 2. In \cite{Li}, Li
found a (weak) DT of the EE \eqref{e} with the Lax pair
\eqref{Lax11} and \eqref{Lax12} for a \em zero \rm spectral
parameter. In \cite{LTJH}, the weak DT was extended to a general \em
nonzero \rm spectral parameter and many kinds of exact solutions
including the solitary, Rossby, conoid, and Bessel waves were
obtained subsquently.

Here we derive the weak DT theorem directly from the general group
Theorem 2.

{\em Theorem 5 (Weak DT theorem).} If $\{\omega',\ \psi',\ \phi'\}$
is a solution of the (2+1)-dimensional EE \eqref{e} and its Lax pair
\eqref{Lax11} and \eqref{Lax12} with the spectral parameter
$\lambda'$, $g(f)$ being an arbitrary function of $f$ which is a
given spectral function of \eqref{Lax11} and \eqref{Lax12} under the
spectral parameter $\lambda_0$, then
\begin{eqnarray}
\{\omega,\ \psi,\ \phi\}=\{\omega'+q,\ \psi'+p,\ \exp(g)\phi'\}
\label{New}
\end{eqnarray}
with the spectral parameter $\lambda$ is a solution of the weak Lax
pair \eqref{Lax11} and \eqref{Lax12} and then Eq. \eqref{cdot} where
$p$ and $q$ are determined by
\begin{eqnarray}
&&q= p_{xx}+p_{yy}\label{q0}\\
&&[p,\ \ln\phi]=0,\label{p0}\\
&&[q,\ \ln\phi']+\lambda'-\lambda+\lambda_0fg_f=0.\label{pq}
\end{eqnarray}

{\em Proof.} Known from the proof of the group Theorem 2, Eqs.
\eqref{det1} and \eqref{det2} are equivalent to Eqs. \eqref{e12a}
and \eqref{e12}. Taking $\xi=x,\ \eta=y$ and $\tau=t$ in Eqs.
\eqref{e12a} and \eqref{e12}, we have
$$ [\omega,\ \phi']+([\omega,\ g]-\lambda)\phi'=0,\eqno(20')$$
$$ [\psi,\ \phi']+\phi'_t+(g_t+[\psi,\ g])\phi'=0. \eqno(21')$$
Substituting Eq. \eqref{New} into Eq. (20'), we have
\begin{eqnarray*}
&&[\omega',\ \phi']+[q,\ \phi']+([\omega',\ g]+[q,\
g]-\lambda)\phi'\\
&&\quad=\lambda' \phi'  +[q,\ \phi'] + g_f\lambda_0 f \phi'+([q,\
g]-\lambda)\phi'\\
&&\quad=\left\{\lambda'-\lambda+\lambda_0 f g_f + [q,\ \ln
\phi']\right\}\phi'=0.
\end{eqnarray*}
Equation \eqref{pq} is proven.

Substituting Eq. \eqref{New} into Eq. (21') yields
\begin{eqnarray*}
&&[\psi',\ \phi']+[p,\ \phi']+\phi'_t+(g_t+[\psi',\ g]+[p,\
 g])\phi'\\
 &&\quad = [p,\ \phi']+[p,\ g]\phi'= [p,\ \ln\phi'+g]\phi'= [p,\ \ln\phi]\phi'=0.
\end{eqnarray*}
Equation \eqref{p0} is proven, and Eq. \eqref{q0} is a direct result
from the definition equation of the vorticity. Theorem 5 is proven.
$\Box$

It is interesting that if all the parameters $\lambda,\ \lambda_0$
and $\lambda'$ are zero and the vorticity of the seed solution is
not a constant, the B\"acklund transformation
$$\omega=\omega'+q,\ \psi= \psi'+p$$ with Eqs. \eqref{p0} and \eqref{pq} is equivalent
to what were obtained by Li \cite{Li}. To see it more clearly, one
can write Eqs. \eqref{p0} and \eqref{pq} in the alternative forms by
eliminating $\phi'_y$ via the Lax pair \eqref{Lax11} and
\eqref{Lax12},
\begin{eqnarray}
&&\lambda'\omega'_x-[\lambda+\lambda_0fg_f](\omega'+q)_x+[\omega',\ q][\ln \phi]_x=0,\label{pq0}\\
&&[\omega'+q,\ p][\ln\phi]_x+\lambda'p_x=0.\label{pq1}
\end{eqnarray}
The equivalent forms of Eqs. \eqref{pq0} and \eqref{pq1} can also be
obtained directly from Eqs. \eqref{det1} and \eqref{det2} by setting
$\xi=x,\ \eta=y$ and $\tau=t$.

{\em Remark 5.} If the seed solution has a constant vorticity, the
equation systems \eqref{p0}--\eqref{pq1} are completely not
equivalent. Actually, when one takes a constant vorticity as a seed
for the zero spectral parameters, nothing can be obtained from Eqs.
\eqref{pq0} and \eqref{pq1}. However, one can really find some
nontrivial solutions from Eq. \eqref{pq} with a constant vorticity
seed. In \cite{LouLi}, the weak DT theorem has been used to obtain
some types of exact solutions such as the solitary waves, the conoid
periodic waves, the Rossby waves, and many kinds of Bessel waves.
Here we will not discuss them further.

\section{Summary and discussion.}

The analytical and exact forms of the vortices and circumfluence of
the two-dimensional fluid are studied by means of the general
symmetry group Theorem 2 of the (2+1)-dimensional EE. Some solution
theorems for the (2+1)-dimensional EE are obtained from the group
theorem by taking special seed solutions. A special weak DT of the
(2+1)-dimensional EE is also obtained from the general group
theorem.

The special solution Theorem 4 gives a quite general exact explicit
solution which covers many kinds of possible vortices and
circumfluence such as the lump-type vortices, dromion-type vortices,
ring solitons, etc. The vortex and circumfluence solutions may have
applications in various physical fields mentioned in the
Introduction and Refs. \cite{fluid}--\cite{H}. Particularly, they
can qualitatively explain some fundamental problems of TCs such as
their eye, track, and the relation between the track and the
background wind, and the relation can be used to predict well the TC
tracks. Hurricanes and/or typhoons have tremendously and
increasingly caused destruction of our world. The method in this
paper provides a possible way to understand and study similar
disasters intensively. As an original study in this aspect, some
introductory analysis by means of our method of Hurricane Katrina
2005 (which almost completely destroyed a whole city, New Orleans)
are presented. The technological observations and phenomenological
discussions on Hurricane Katrina 2005 can be found in many papers
\cite{Katrina}. In this paper, an approximate analytical expression
for the (2+1)-dimensional stream function of Katrina 2005 is
obtained. The expression is an exact solution of the
(2+1)-dimensional EE and includes some messages including the eye
size, the hurricane size, the strength, the relation between the
hurricane center and the steering flow, etc. The relation is also
used to predict the track of the hurricane.

The discovery of the general group theorem may lead to the discovery
of various interesting exact solutions which can be applied to many
real physical fields. This paper is just the beginning study in this
aspect. There are various important problems should be studied
further. For instance, the possible solutions from the general group
Theorem 2 are only discussed in three very special cases: (i) the
constant vorticity seed (Corollary 1), (ii) the zero spectral
parameters without the gauge transformation (Corollary 2, Theorems 3
and 4), and (iii) the pure weak DT case (Theorem 5).

In this paper, we only discuss the (2+1)-dimensional EE. Two types
of Lax pairs of the (3+1)-dimensional EE have also been given in
\cite{Li1} and some special DTs of these Lax pairs have also been
given in \cite{LTJH}. However, these DTs have not yet been
utilized to find exact solutions of the (3+1)-dimensional EE.
Furthermore, the corresponding symmetry groups similar to that of
the (2+1)-dimensional EE given in this paper have not yet been
discussed.

The Lax pair and then the DT found in this paper have only weak
meaning. Whether the (2+1)-dimensional EE is integrable under some
stronger meanings [similar to those of (3+1)-dimensional EE] is
still open.

The more general applications of the vortex solutions given in this
paper both in atmospheric dynamics and in other physical fields
deserve more investigations. Especially, to describe the hurricane
more effectively and accurately, some other important factors such
as the Coriolis force and the viscosity of the fluid must be
considered.

Because of the importance of the EEs and the NS system and their
wide applications, the models and all the problems mentioned above
are worthy of further study.

\begin{acknowledgments}
The authors are grateful for helpful discussions with Professor. Y.
S. Li, Professor D. H. Luo, Professor Y. Chen, Professor X. B. Hu,
and Professor Q. P. Liu. This work was supported by the National
Natural Science Foundation of China (Grants No. 10475055, No.
40305009, No. 90503006, and No. 10547124), Program for New Century
Excellent Talents in University (NCET-05-0591), Shanghai
Post-doctoral Foundation (06R214139), Shandong Taishan Scholar
Foundation and National Basic Research Program of China (973
program) (Grant No. 2005CB422301).

\end{acknowledgments}


\begin{thebibliography}{999}
\bibitem{NS}D. Sundkvist, V. Krasnoselskikh, P. K. Shukla, A. Vaivads, M. Andr\'e, S. Buchert and H. R\`eme, Nature,
\bf 436  \rm  825 (2005); G. Pedrizzetti, Phys. Rev. Lett. \bf 94
 \rm  194502 (2005).
\bibitem{clay} C. L. Fefferman,
http://www.claymath.org /millennium /Navier-Stokes \_Equations
/Official\_Problem\_Description. pdf (2000).
\bibitem{Li}Y. G. Li, J. Math. Phys. \bf 42 \rm  3552 (2001).
\bibitem{Li1}
Y. G. Li and A. V. Yurov, Stud. Appl. Math. \bf 111  \rm  101
(2003).
\bibitem{LTJH}S. Y. Lou, X. Y. Tang, M. Jia and F. Huang, \em
Vortices, circumfluence, symmetry groups and Darboux
transformations of the Euler equations, \rm nlin.PS/0509039.
\bibitem{fluid}
P. H. Chavanis and J. Sommeria, Phys. Rev. Lett. \bf 78 \rm   3302
(1997); P. H. Chavanis, \em ibid  \bf 84 \rm 5512 (2000).
\bibitem{plasma}
E. Cafaro, D. Grasso, F. Pegoraro, F. Porcelli and A. Saluzzi,
Phys. Rev. Lett. \bf 80 \rm   4430 (1998); D. Del Sarto, F.
Califano and F. Pegoraro, \em ibid \bf 91 \rm 235001 (2003).
\bibitem{ocean}
V. M. Canuto and M. S. Dubovikov, Ocean Modelling \bf 8
 \rm  1 (2005).
\bibitem{gas}C. Girard, R. Benoit, M. Desgagne,  Monthly Weather Rev. \bf 133 \rm   1463 (2005);
S. Kurien, V. S. L'vov, I. Procaccia and K. R. Sreenivasan, Phys.
Rev. E \bf 61 \rm   407 (2000).
\bibitem{super}F. D. M. Haldane and Y. S. Wu, Phys. Rev. Lett. \bf 55 \rm   2887 (1985).
\bibitem{astro}
S. Bonazzola, E. Gourgoulhon and J. A. Marck, Phys. Rev. D \bf 56
\rm   7740 (1997). 
\bibitem{sta}A. J. Niemi, Phys. Rev. Lett. \bf 94 \  \rm   124502
(2005).
\bibitem{particle}L.Faddeev, A.J.Niemi, and U.Wiedner, hep-ph/0308240.
\bibitem{bose}A. J. Leggett, Rev. Mod. Phys. \bf 73 \rm   307 (2001).
\bibitem{crystal}I. Chuang, R. Durrer, N. Turok, and B. Yurke, Science \bf 251 \rm   1336 (1991);
M. J. Bowick, L. Chandler, E.A. Schiff, and A. M. Srivastava,
Science \bf 263 \rm   943 (1994).
\bibitem{H}E. Babaev, A. Sudb\o, and N.W.
Ashcroft, Nature (London) \bf 431  \rm  666 (2004).
\bibitem{Lamb}H. Lamb, \em Hydrodynamics, \rm 6th ed. (Dover, New York, 1945).
\bibitem{Sov}A. A. Abrashkin and E. I. Yakubovich, Sov. Phys. Dokl.
 \bf 276 \rm 370 (1984).
\bibitem{Yu}
A. V. Yurov and A. A. Yurova, Theor. Math. Phys., \bf 147 \rm 501
(2006).
\bibitem{Jia}S. Y. Lou, M. Jia, F. Huang and X. Y. Tang, Int. J.
Theor. Phys. (2007) DOI:10.1007/s10773-006-9327-5.
\bibitem{HF}F. Huang and S. Y. Lou, Phys. Lett. A \bf 320 \rm    428
(2004); V. L. Saveliev, M. A. Gorokhovski, Phys. Rev. E \bf 72 \rm
016302 (2005).
\bibitem{group_CSF}S. Y. Lou and H. C. Ma, Chaos, Solition \& Fractals \bf 30 \rm 804 (2006).
\bibitem{group_JPA}S. Y. Lou and H. C. Ma, J. Phys. A: Math. Gen. \bf
38 \rm L129 (2005), S. Y. Lou, Chin. Phys. Lett.  \bf 21 \rm 1020 (2004).
\bibitem{ring}S. Y. Lou, J. Math. Phys. \bf 41 \rm   6509 (2000).
\bibitem{ring1}X. Y. Tang, S. Y. Lou and Y. Zhang, Phys. Rev. E. \bf 66 \rm   046601
(2002).
\bibitem{typhoon}K. Emanuel, Nature, \bf 436 \rm   686 (2005).
\bibitem{eye}Q. H. Zhang, S. J. Chen, Y. H. Kuo and R. A. Anthes,
Monthly Weather Rev. \bf 133 \rm   725 (2005);
\bibitem{noaa}
http://www.katrina.noaa.gov/satellite/satellite.html.
\bibitem{G}http://fermi.jhuapl.edu/hurr/05/katrina/katrina.txt.
\bibitem{track}
http://www.ncdc.noaa.gov/img/climate/research/2005/katrina/katrina.gif.
\bibitem{G1}http://weather.unisys.com/hurricane/atlantic/2005H/KATRINA/track.dat.
\bibitem{Katrina}A. Apple, Nature \bf 437  \rm   462 (2005);
R. Dalton, \it ibid   \rm  300; E. Stokstad, Science  \bf 310  \rm
 1264 (2005); J. Kaiser, \it ibid  \rm  1267 (2005).
\bibitem{JiaLou} M. Jia, C. Lou and S. Y. Lou, Chin. Phys. Lett. \bf 23 \rm 2878 (2006).
\bibitem{LouLi}S. Y. Lou and Y. S. Li, Chin. Phys. Lett. \bf 23 \rm 2633 (2006).
\end{thebibliography}
\end{document}